\documentclass{article}
\usepackage{arxiv}
\usepackage[utf8]{inputenc} 
\usepackage[T1]{fontenc}    
\usepackage{hyperref}       
\usepackage{url}            
\usepackage{booktabs}       
\usepackage{amsfonts}       
\usepackage{nicefrac}       
\usepackage{microtype}      
\usepackage{lipsum}
\usepackage{graphicx}
\usepackage[T1,OT1]{fontenc}
\usepackage{braket}
\usepackage{soul}
\usepackage{multirow}
\usepackage{graphicx}%
\usepackage{dcolumn}%
\usepackage{bm}%
\usepackage{caption}
\usepackage{amsmath}
\usepackage{amssymb}
\usepackage[T1]{fontenc}
\usepackage[utf8]{inputenc}   
\usepackage{lmodern}

\usepackage[labelformat=parens]{subcaption}
\usepackage{float}
\usepackage{romannum}
\usepackage{xcolor}

\usepackage{cite}

\title{The Market Crystal: A Spin-Lattice Model for Collective Cryptocurrency States}

\author{
 Hamidreza Oliaei-Moghadam \\
  Department of Physics, College of Sciences\\
  Shiraz University\\
  Shiraz 71946-84795, Iran \\
  \texttt{oliaeimoghaddam@gmail.com}
}
 
\begin{document}
\maketitle

\noindent
\begin{abstract}
	Collective dynamics in financial markets can emerge through synchronized movements of large groups of assets. Motivated by analogies with interacting many-body systems, we introduce a spin-lattice representation for analyzing collective states in cryptocurrency markets. In this framework, assets are encoded as binary spin variables according to the sign of their returns, while correlations between assets determine effective interaction strengths. A correlation-based breadth-first search (CBFS) procedure embeds 169 cryptocurrencies into a $13 \times 13$ lattice, enabling the construction of an Ising-like Hamiltonian describing the market configuration, which we call the \emph{Market Crystal}. Macroscopic observables such as magnetization and energy provide a statistical-mechanical characterization of collective market states. The resulting phase-space structure highlights regimes of strong alignment and fragmentation among assets, with an energy--magnetization pattern suggestive of predominantly ferromagnetic interactions. This framework offers a statistical-mechanical viewpoint for studying collective behavior in financial systems.
\end{abstract}


\section{Introduction}

Financial markets are complex systems composed of many interacting assets whose price movements exhibit significant statistical dependencies. These interactions can generate collective dynamics that manifest as large-scale patterns and coordinated movements across the market. Periods of expansion, contraction, and financial stress often involve the simultaneous movement of many assets, indicating that market behavior cannot always be understood solely through the analysis of individual price series. Instead, system-level organization emerges from interactions among assets, giving rise to collective trends, synchronization, and regime shifts \cite{cont2000,gopikrishnan2001,pan2007,munnix2012}. This field of research, often termed econophysics, has established that complex market dynamics can be analyzed using foundational concepts from statistical physics and financial modeling \cite{mccauley2004,sinha2010,chakraborti2011,cipra1987}.

The presence of such collective phenomena has motivated the use of methods from statistical physics to study financial markets. In particular, spin-based and Ising-type models provide a simple framework for describing systems made of many interacting units that can exhibit collective behavior. These models have been applied in many areas, including socio-economic dynamics, biological and physical systems, as well as imaging applications such as single-pixel reconstruction \cite{stanley2008,zhou2007,oliaei2023,oliaei2025,singh2020,lipowski2022}. Furthermore, specific spin-based models have been developed to capture the micro-to-macro dynamics in financial systems, aiding the understanding of collective market transitions and punctuated equilibrium \cite{ponzi2000,kaizoji2002,krause2013}. One advantage of these models is that they connect interactions at the microscopic level with observable quantities at the system level, such as magnetization and energy, which can be used to describe collective states.

Approaches inspired by many-body systems have been used to study the structure and dynamics of financial markets, showing that markets often display properties similar to interacting physical systems. Empirical studies have found that asset returns exhibit nontrivial correlation structures, indicating that assets do not move independently but are influenced by common market factors and interactions within the market \cite{plerou2002,stanley2008,laloux2000,voit2005}.

Several studies have used spin-based models from statistical mechanics to study financial market dynamics. In many of these models, binary spin variables represent the decisions or expectations of interacting agents. This approach makes it possible to study phenomena such as herding behavior, market polarization, and collective fluctuations \cite{zhou2007,eckrot2016,lima2017,cividino2023}. These agent-based models help explain how local interactions among traders can lead to macroscopic market behavior \cite{krawiecki2005,lux1999}. Related work has also considered binary representations of asset price movements and pairwise interaction models based on maximum-entropy principles \cite{borysov2015}.

At the same time, empirical studies of financial markets have emphasized the importance of correlation structures among assets. Correlation matrices and network-based representations have been widely used to characterize the organization of financial markets, revealing hierarchical structures, sector clustering, and dominant market modes \cite{mantegna1999,onnela2003trees,plerou2002,laloux2000}. These representations, often supported by the application of random matrix theory and topological filtering, provide a robust way to analyze the evolution of dependencies and hierarchical organization in financial time series \cite{bouchaud2009,onnela2003dynamic,tumminello2007,fenn2011,marti2021}. These approaches provide valuable insight into relationships between assets, but they mainly describe network structure and do not directly provide a Hamiltonian framework that allows the study of thermodynamic quantities.

Despite these advances, a gap remains between correlation-based descriptions of asset interactions and spin-based models used to study collective dynamics. Correlation studies capture the interaction structure between assets, but they usually do not provide a statistical-mechanical framework for defining energies and macroscopic observables for market configurations. At the same time, many spin models focus on the behavior of agents rather than representing the configuration of assets themselves \cite{voit2005,stanley2008,zhou2007}. A framework that combines asset correlations with a spatially organized spin representation can therefore provide a useful way to study collective market states.

Cryptocurrency markets provide a suitable environment for exploring such representations. Compared with traditional financial markets, cryptocurrencies often show strong collective movements and noticeable co-movement among assets, frequently influenced by dominant market factors and large leading assets such as Bitcoin. Several studies have reported significant correlation structures and collective modes in cryptocurrency markets \cite{watorek2021,borri2022,aslanidis2019}, suggesting that these markets may be well suited for analysis using simplified statistical-mechanical models.

In this work, we introduce a spin-lattice framework for representing collective states in cryptocurrency markets. In this approach, individual assets are represented as binary spin variables determined by the sign of their returns, while empirical correlations between assets define effective interaction strengths. The assets are placed on a lattice structure that allows the construction of an Ising-type Hamiltonian describing the compatibility between instantaneous market configurations and the underlying correlation structure \cite{voit2005,stanley2008}. Within this framework, macroscopic observables such as magnetization and energy serve as system-level indicators of market alignment and collective organization.

The remainder of this paper is organized as follows. Section~\ref{subsec:data_spins} describes the dataset and the encoding of asset returns into spin variables. Section~\ref{subsec:embedding_lattice} explains the correlation-based procedure used to place assets on the lattice and construct the interaction network. Section~\ref{subsec:hamiltonian_observables} presents the Hamiltonian formulation and the macroscopic observables used to characterize collective market states. Section~\ref{result_discussion} presents the empirical results of the Market Crystal framework. Finally, Section~\ref{sec:conclusion} concludes the paper.

\section{The Market Crystal: A Spin-Lattice Framework}
\label{sec:model_construction}

\subsection{Data, Returns, and Spin Encoding}
\label{subsec:data_spins}

The empirical foundation of this study relies on historical price data for a universe of $N = 169$ cryptocurrency assets. The assets were selected primarily based on their market capitalization and the availability of continuous historical price data over the study period, ensuring that the sample represents actively traded cryptocurrencies with sufficient market activity. The dataset spans the period from September 10, 2022 to November 30, 2025, corresponding to 1178 daily observations. For each asset $i$, the price time series $P_i(t)$ is recorded at daily frequency. Daily closing prices for USDT trading pairs were obtained from the Binance public API. From the price time series we compute logarithmic returns \cite{voit2005,cont2000}.

For each asset $i$, the return at time $t$ is defined as
\begin{equation}
	r_i(t) = \ln P_i(t) - \ln P_i(t-1),
	\label{return}
\end{equation}
where $P_i(t)$ denotes the daily closing price of asset $i$ at time $t$.

To construct a discrete statistical-mechanical representation, the return dynamics are coarse-grained into binary spin variables. For each asset $i$ and time $t$, the spin state is defined as

\begin{equation}
	\sigma_i(t)=
	\begin{cases}
		+1, & r_i(t) > 0,\\
		-1, & r_i(t) < 0,\\
		\pm 1 \text{ with equal probability}, & r_i(t)=0.
	\end{cases}
	\label{return_random}
\end{equation}

This mapping transforms the multivariate return series into a binary spin configuration analogous to an Ising system used in statistical physics models of financial markets \cite{voit2005,zhou2007,stanley2008}.
At each time $t$, the collective state of the market is represented by the configuration of all spin variables. We denote the instantaneous market configuration as
\begin{equation}
	\boldsymbol{\sigma}(t) = \left(\sigma_1(t), \sigma_2(t), \ldots, \sigma_N(t)\right),
\end{equation}
where $N$ is the total number of assets and each component $\sigma_i(t) \in \{+1,-1\}$ represents the spin state associated with asset $i$. This vector provides a complete binary representation of the market state at time $t$ within the spin-lattice framework.

\subsection{Correlation-Based Embedding and Lattice Construction}
\label{subsec:embedding_lattice}

Unlike particles in a physical crystal, financial assets do not possess an intrinsic spatial arrangement. Instead, relationships between assets are inferred from statistical dependencies between their return time series, typically quantified through correlation \cite{mantegna1999,plerou2002,laloux2000}. The spatial assignment of assets is determined using the empirical Pearson correlation matrix $\rho_{ij}$ computed from the return time series \cite{plerou2002,mantegna1999}. The Pearson correlation coefficient between assets $i$ and $j$ is defined as

\begin{equation}
	\rho_{ij} =
	\frac{\langle r_i r_j \rangle - \langle r_i \rangle \langle r_j \rangle}
	{s_i s_j},
\end{equation}

where $\langle \cdot \rangle$ denotes the time average over the observation period and $s_i$ is the standard deviation of returns for asset $i$. In this study, $\rho_{ij}$ is estimated over the full observation period and treated as a fixed interaction matrix used to evaluate the sequence of observed market configurations.

To construct the Market Crystal, we arrange the $N = 169$ assets on a two-dimensional $13 \times 13$ lattice with open boundary conditions. The lattice provides a convenient spatial structure in which nearby sites represent assets with stronger statistical relationships. The embedding of assets into the lattice is performed through a correlation-based breadth-first search (CBFS) procedure. The process begins by selecting a seed asset that serves as the center of the lattice. In this study, Bitcoin (BTCUSDT) is chosen as the seed due to its dominant role within the cryptocurrency market and its strong correlations with many other assets. Starting from the seed node, assets are iteratively assigned to neighboring lattice sites according to their correlation strength with already-placed assets. At each step, the algorithm identifies assets that exhibit the strongest correlations with the current frontier of embedded nodes and places them into the nearest available lattice positions.

\begin{figure}[htbp]
	\centering
	\includegraphics[width=1\linewidth]{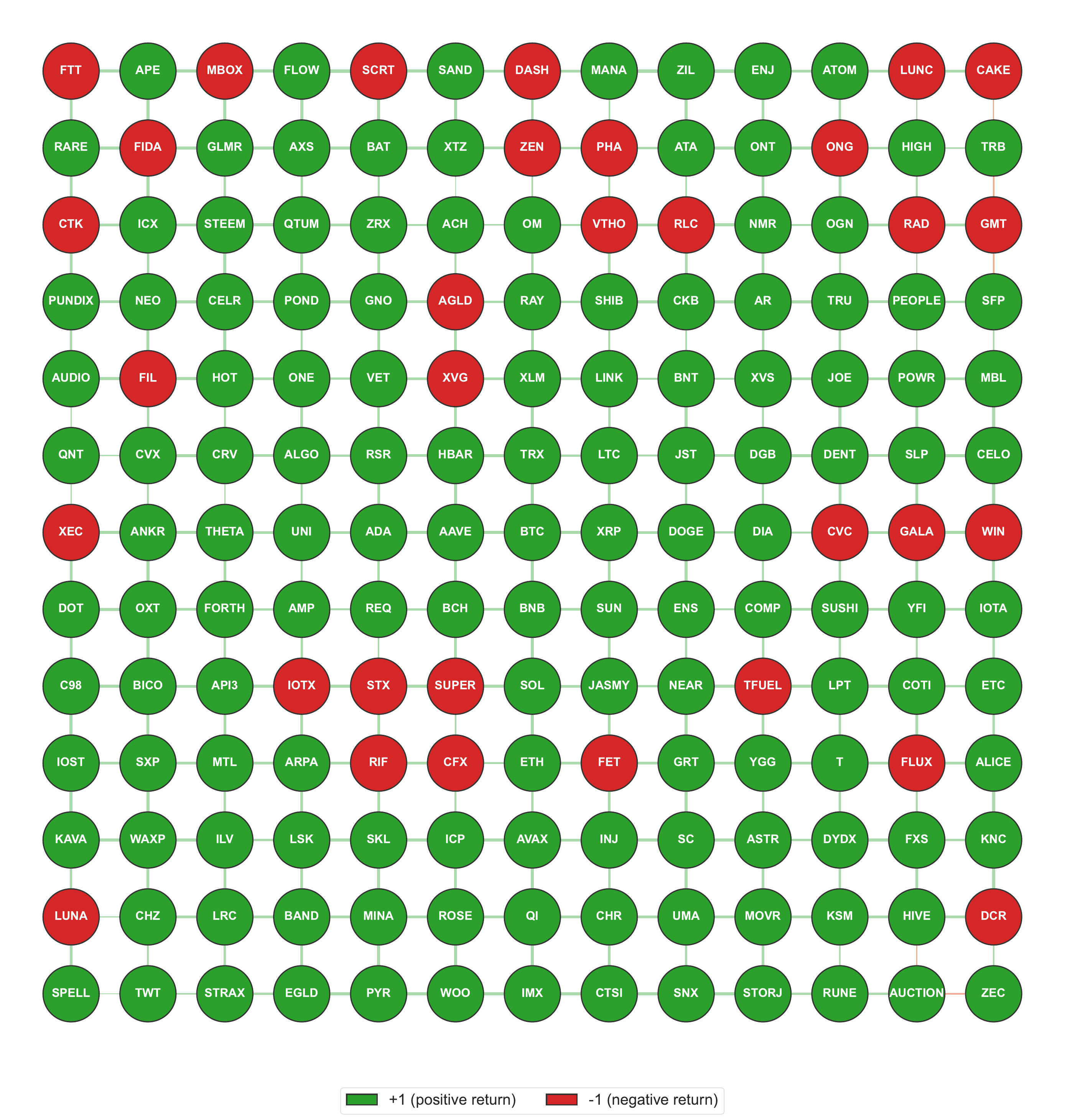}
	\caption{
		Two-dimensional snapshot of the Market Crystal configuration $\boldsymbol{\sigma}(t)$ on May 5, 2023. 
		The $N=169$ assets are arranged on a $13\times13$ lattice obtained through the 
		correlation-based breadth-first search (CBFS) procedure, with Bitcoin (BTC) 
		serving as the central seed node. Node colors represent the instantaneous 
		spin state $\sigma_i(t)$, where green indicates a positive daily return 
		($\sigma_i=+1$) and red indicates a negative return ($\sigma_i=-1$). 
		Edges connect nearest-neighbor lattice sites and represent interaction 
		couplings $J_{ij}=\rho_{ij}$ derived from the empirical Pearson correlation 
		matrix. The configuration shown illustrates a period of strong positive 
		alignment across the market.
	}
	\label{fig:lattice_embedding}
\end{figure}

The resulting lattice embedding provides a two-dimensional representation of the high-dimensional correlation structure of the cryptocurrency market. Figure~\ref{fig:lattice_embedding} illustrates this framework by showing a snapshot of the market configuration on May 5, 2023, where assets are positioned according to their correlation-based embedding and colored according to their instantaneous spin states. Assets with strong positive correlations tend to occupy neighboring lattice sites, while weakly correlated or negatively correlated assets are placed farther apart. The CBFS embedding provides a heuristic spatial representation of the correlation structure rather than a unique optimal configuration. Small variations in the placement of highly correlated assets lead to similar local neighborhoods, and the resulting statistical observables of the system remain qualitatively unchanged.

Once the embedding is complete, the lattice topology defines a network of nearest-neighbor interactions. The coupling strength between adjacent assets $i$ and $j$ is defined directly from their empirical Pearson correlation

\begin{equation}
	J_{ij} = \rho_{ij}.
\end{equation}

Positive correlations produce ferromagnetic couplings that favor spin alignment, whereas negative correlations generate antiferromagnetic interactions that promote opposing spin states \cite{voit2005,stanley2008}.

\subsection{Hamiltonian and Macroscopic Observables}
\label{subsec:hamiltonian_observables}

Given the lattice geometry and interaction strengths, the instantaneous state of the market is characterized using an Ising-like Hamiltonian

\begin{equation}
	H(t) = - \sum_{\langle i,j \rangle} J_{ij}\sigma_i(t)\sigma_j(t),
\end{equation}

where the sum runs over all nearest-neighbor pairs on the lattice \cite{voit2005,cipra1987,stanley2008}. In this framework, the Hamiltonian serves as a descriptive energy functional that characterizes the observed market configuration rather than a dynamical model governing its evolution.

Configurations in which correlated assets move in compatible directions produce lower energy states, while configurations that contradict the empirical correlation structure lead to higher energy values \cite{borysov2015}.

To characterize the collective dynamics of the system, we focus on two macroscopic observables derived from the lattice configuration.

\paragraph{Market Magnetization}

The magnetization is defined as

\begin{equation}
	M(t) = \frac{1}{N}\sum_{i=1}^{N}\sigma_i(t),
\end{equation}

which measures the net alignment of asset movements across the entire market. Values of $M(t)$ close to $+1$ correspond to states in which nearly all asset spins are positive (indicating widespread positive returns), while values near $-1$ indicate that most spins are negative (corresponding to broad market declines). When $M(t)$ is close to zero, positive and negative asset movements are approximately balanced \cite{voit2005,lux1999,zhou2007}.

\paragraph{Energy per Site}

To facilitate comparison across time, we consider the energy per site

\begin{equation}
	E(t)=\frac{H(t)}{N},
\end{equation}

which provides a normalized measure of how well the instantaneous market configuration aligns with the underlying correlation structure. Deep negative values correspond to configurations in which strongly correlated assets move in mutually compatible directions, while higher values indicate increasing levels of frustration or inconsistency with the correlation structure \cite{stanley2008,plerou2002}.

\section{Results and Discussion}
\label{result_discussion}

Having defined the macroscopic observables of the Market Crystal, we now examine their empirical behavior across the cryptocurrency market over the full observation period. The analysis focuses on the temporal evolution of market magnetization and system energy, as well as the joint phase-space structure formed by these quantities. Together, these observables reveal how collective market states emerge, persist, and transition over time.

\subsection{Magnetization Dynamics}
We first examine the temporal evolution of the market magnetization $M(t)$, which measures the degree of directional consensus among assets. Values of $M(t)$ close to $+1$ or $-1$ correspond to strongly ordered market states in which most assets move in the same direction, whereas values near zero indicate fragmented behavior with no dominant market-wide trend.
To highlight medium-term collective behavior, we also compute a 21-day moving average of the magnetization.
\begin{figure}[htbp]
	\centering
	\includegraphics[width=1\linewidth]{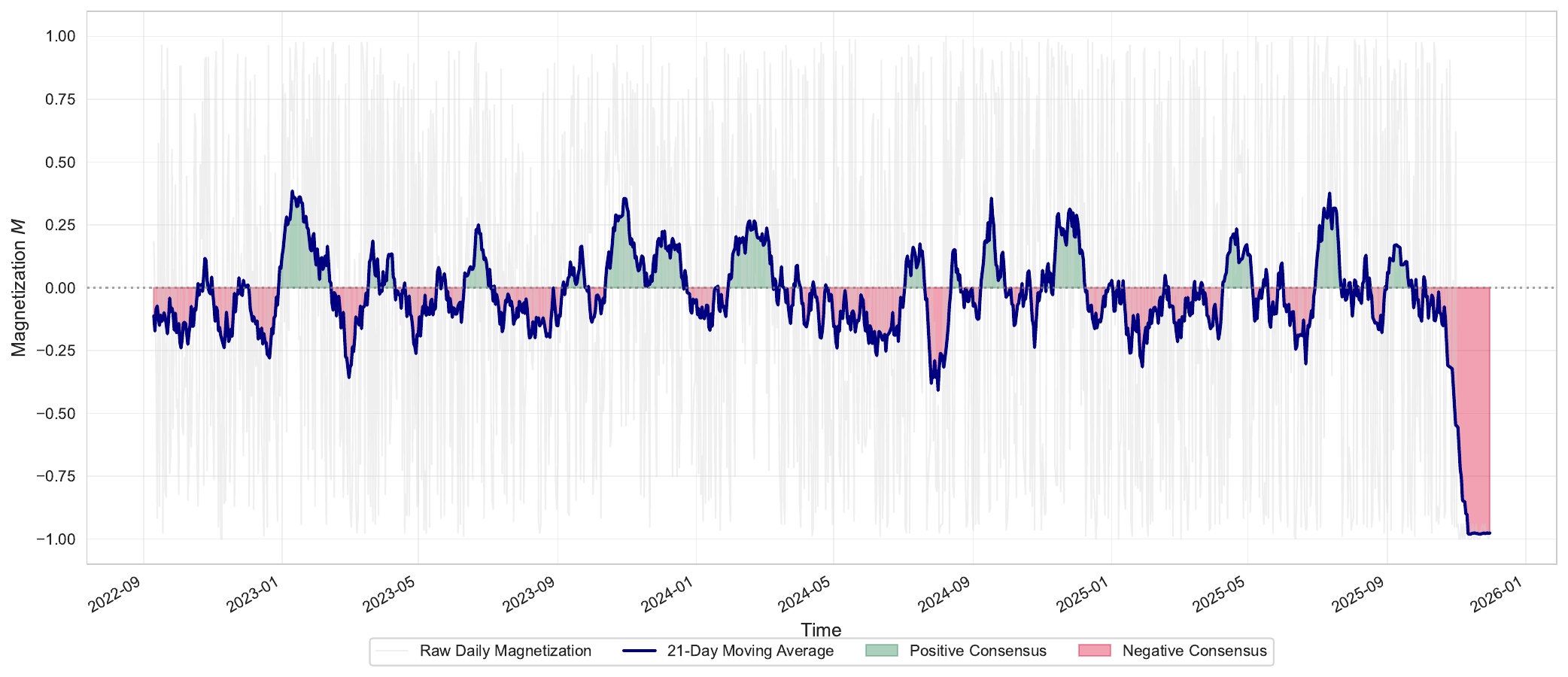}
	\caption{Temporal evolution of the market magnetization $M(t)$. The faint gray curve shows the raw daily series, while the solid black curve denotes the 21-day moving average. Green and red shaded regions mark positive and negative ordered regimes, respectively, in which the market exhibits strong collective alignment. Periods where the smoothed magnetization remains near zero correspond to comparatively disordered market states.}
	\label{fig:magnetization}
\end{figure}

As shown in Fig.~\ref{fig:magnetization}, the cryptocurrency market alternates between extended intervals of collective alignment and more disordered regimes. Sustained positive $M$ corresponds to broad upward market movement, while sustained negative $M$ reflects coordinated market decline. In both cases, large deviations from zero indicate pronounced herding behavior across the asset universe.

A notable empirical feature is that the most pronounced bearish episodes coincide with strongly negative magnetization. In the observed sample, values $M \lesssim -0.5$ are typically associated with periods of strong negative collective alignment and coincide with broad market decline. The final segment of the dataset is especially striking, as the smoothed magnetization approaches $M=-1$, corresponding to an extreme ordered state in which downward movement becomes nearly fully synchronized across the lattice. Because the sample ends in this regime, no subsequent recovery is observed within the available time window.

By contrast, intervals in which $M$ remains close to zero are associated with weaker collective coordination and more heterogeneous asset dynamics. The repeated alternation between ordered positive, ordered negative, and disordered regimes suggests that the market undergoes recurrent collective-state transitions, consistent with the statistical-mechanical interpretation of the Market Crystal framework.

\subsection{Energy Evolution}
We next examine the temporal evolution of the energy per site $E(t)$, which reflects how closely the instantaneous market configuration aligns with the empirical correlation structure encoded in the interaction matrix $J_{ij}$.
\begin{figure}[htbp]
	\centering
	\includegraphics[width=1\linewidth]{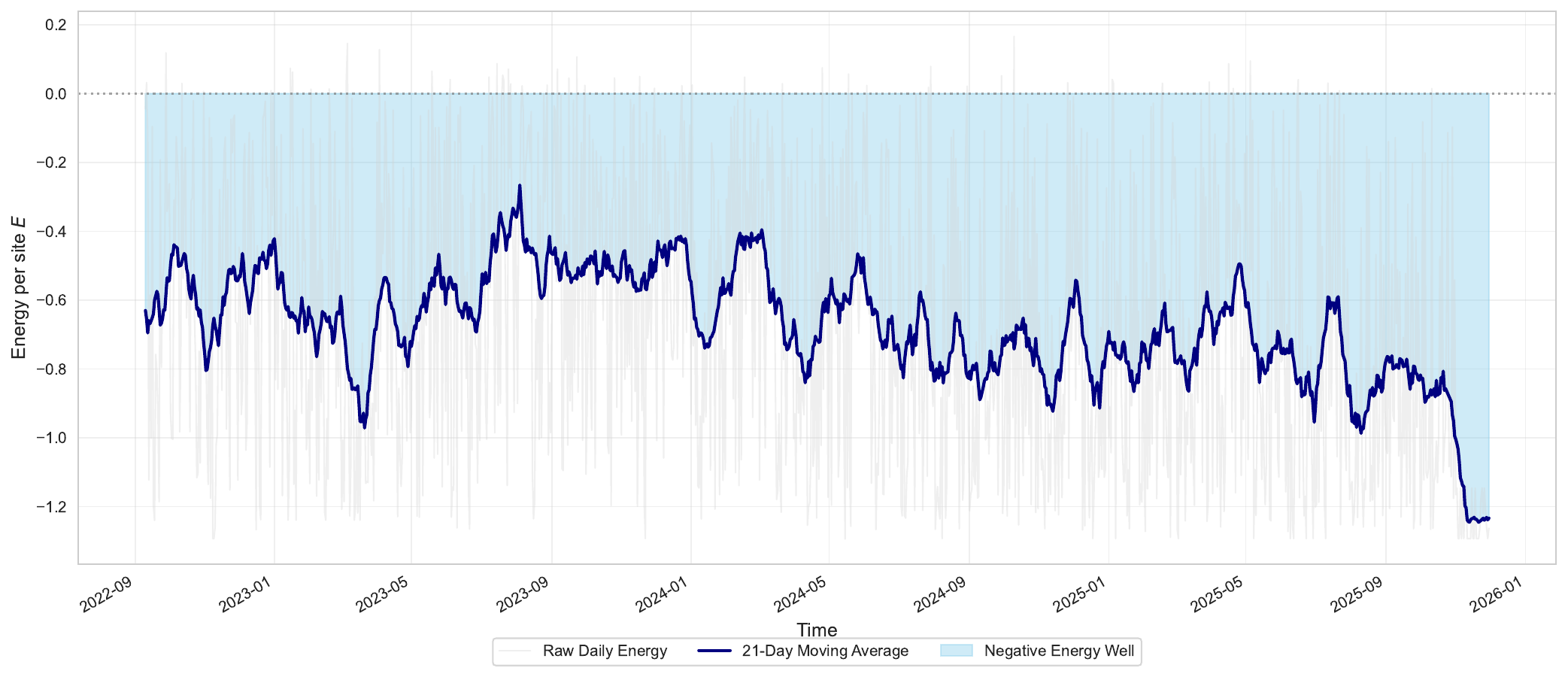}
	\caption{\textbf{Energy dynamics of the Market Crystal.}
		The time series shows the evolution of the system’s energy per site $E(t)$ computed from the empirical interaction matrix $J_{ij}=\rho_{ij}$. Light gray lines represent the raw daily energy values, capturing high-frequency fluctuations of the microscopic configuration. The black curve denotes the 21-day moving average, revealing the slower structural evolution of the market state. The shaded gray region highlights the negative energy well, indicating periods in which asset states align with the empirical correlation structure and the system occupies a coherent configuration. A dashed horizontal line marks the neutral reference level $E=0$. Deeper excursions into the negative energy well correspond to stronger collective alignment among assets, while movements toward zero indicate weakening structural coherence and increased systemic tension.}
	\label{fig:energy}
\end{figure}
Figure~\ref{fig:energy} presents the resulting energy time series. The raw daily values fluctuate substantially, reflecting short-term variations in the microscopic configuration of asset movements. However, the smoothed trajectory reveals a clearer structural pattern. For most of the sample, the system remains in a negative-energy regime, indicating that the observed asset configurations are broadly consistent with the empirical correlation structure.

Deeper excursions into the negative-energy well correspond to periods of stronger synchronization across correlated assets, reflecting highly coherent market states. Conversely, movements toward the zero-energy level indicate partial breakdowns of this coherence, where the instantaneous configuration becomes less consistent with the underlying correlation network. Such episodes can therefore be interpreted as periods of structural tension, in which the market configuration departs from the correlation-driven organization of the system. In the language of spin systems, these states correspond to increased frustration, where the instantaneous configuration cannot simultaneously satisfy the interaction preferences of many neighboring asset pairs.

Viewed together with magnetization, the energy provides a complementary perspective on market dynamics. While magnetization measures the net directional alignment of asset returns, the energy captures the degree to which the observed configuration respects the interaction structure of the system.

\subsection{Energy--Magnetization Phase Space}
\label{subsec:phase_space}

To further characterize the collective behavior of the Market Crystal, we examine the phase space defined by $M(t)$ and $E(t)$. This representation provides a compact physical description of the market by combining a measure of global directional alignment with a measure of structural coherence.

\begin{figure}[htbp]
	\centering
	\includegraphics[width=0.90\linewidth]{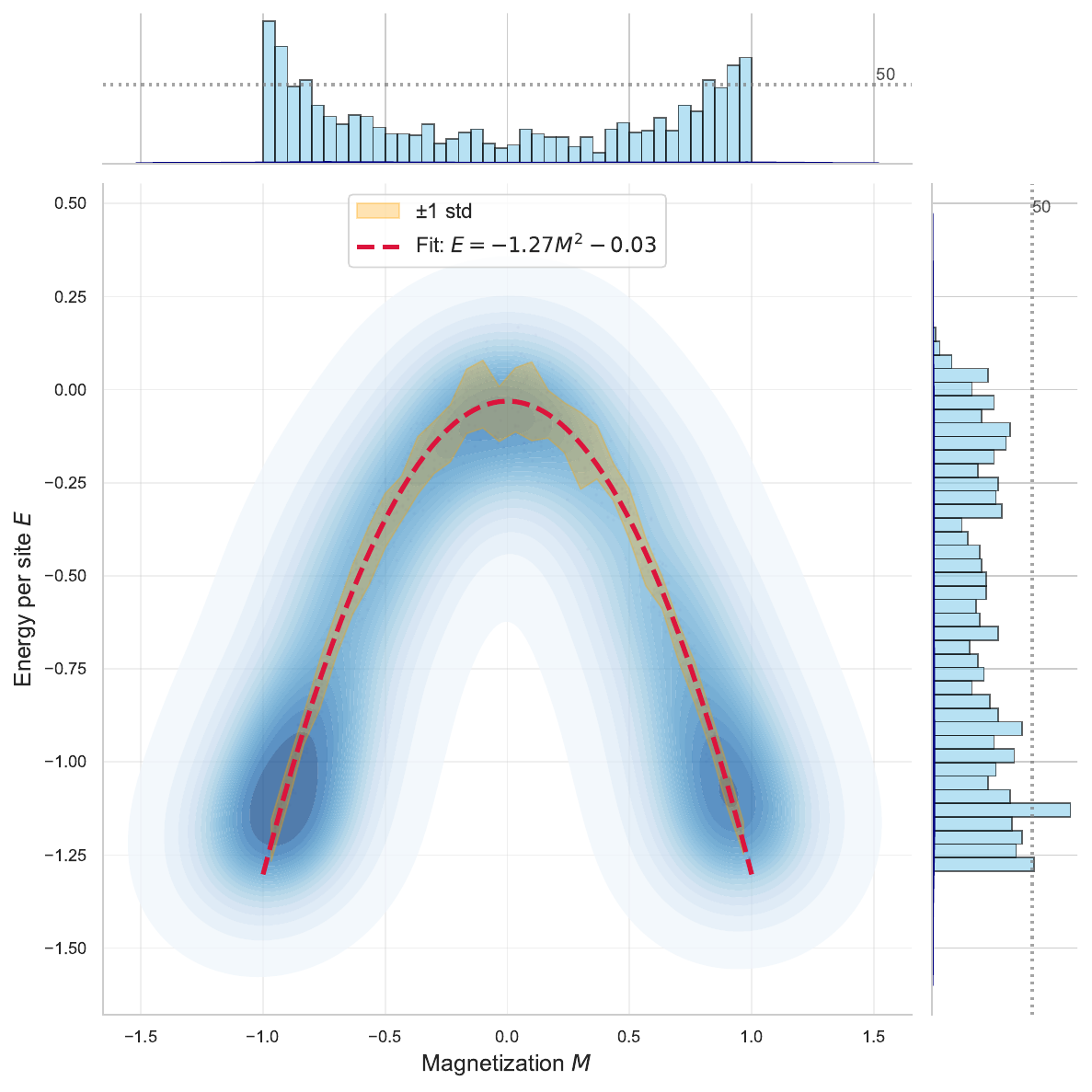}
	\caption{\textbf{Energy--magnetization phase space of the Market Crystal.}
		The blue density map shows the empirical distribution of market states in the $(M,E)$ plane. The marginal histograms display the separate distributions of magnetization and energy; notably, the magnetization histogram exhibits a pronounced bimodal structure. The dashed red curve is a quadratic fit, approximately given by $E=-1.27M^2-0.03$, and the orange band indicates the spread of energy values at different magnetization levels. The distribution follows a clear parabolic structure, with lower energies observed at larger $|M|$ and a broader energy range near $M\approx0$.}
	\label{fig:phase_space}
\end{figure}

As shown in Fig.~\ref{fig:phase_space}, the phase space is organized around a simple curved structure that is approximately described by
\begin{equation}
	E \simeq -1.27M^2 - 0.03 .
	\label{eq:phase_fit}
\end{equation}
This relation shows that the energy decreases as the absolute value of the magnetization increases. In other words, market states with stronger overall alignment across assets tend to occupy lower-energy regions. This behavior arises because the Hamiltonian depends on pairwise spin products $\sigma_i\sigma_j$, so configurations with stronger global alignment naturally produce larger numbers of energetically favorable neighboring interactions.

Crucially, the marginal distribution of $M$ (shown in the top histogram of Fig.~\ref{fig:phase_space}) reveals a distinct bimodal structure, with high-frequency peaks located near the extreme ordered states at $M \approx \pm 1$. This indicates that the Market Crystal spends a significant portion of its time in highly synchronized regimes of near-universal growth or decline. Rather than fluctuating around a disordered neutral state ($M \approx 0$), the system exhibits a strong preference for these "ferromagnetic-like" ordered phases.

This pattern reflects the predominantly ferromagnetic nature of the interaction network. Because most neighboring assets on the lattice are positively correlated, configurations in which nearby assets move in the same direction are energetically preferred. The approximately parabolic form therefore arises naturally from the tendency of correlated neighboring assets to align in low-energy configurations. The widest spread of energy appears near $M\approx0$. This means that when the market has little net directional bias, its internal organization can still vary substantially. Some near-neutral states remain relatively coherent and lie at lower energy, while others are more weakly organized and appear closer to zero energy. Thus, similar values of magnetization do not always correspond to the same level of structural order.

At larger positive or negative values of $M$, the energy distribution becomes narrower. Highly aligned regimes therefore occupy more localized regions of phase space, indicating a more constrained and coherent market structure.

A further feature of the density map is the asymmetry between the two low-energy branches. The branch at negative magnetization is more populated than the one at positive magnetization, showing that coherent bearish regimes occurred more often, or lasted longer, than comparable bullish regimes during the sample period.

Overall, the energy--magnetization phase space gives a clear thermodynamic picture of the cryptocurrency market. The bimodal occupancy of the phase space suggests that the market is fundamentally an ordered system that oscillates between extreme collective states, while weakly aligned regimes near $M\approx0$ act as transient or less frequent configurations. This phase-space representation therefore provides a compact thermodynamic description of market organization, where magnetization plays the role of an order parameter and the energy quantifies the compatibility of the instantaneous configuration with the empirical interaction network.

\section{Conclusion}
\label{sec:conclusion}

In this paper, we introduced the ``Market Crystal,'' a spin-lattice framework that maps the high-dimensional dynamics of cryptocurrency markets onto a structured two-dimensional Ising-like system. Using a correlation-based breadth-first search (CBFS) algorithm, we embedded 169 digital assets into a $13 \times 13$ lattice and represented their daily return signs as binary spin variables, thereby transforming complex market behavior into a form that can be analyzed using the language of statistical physics.

This construction makes it possible to study cryptocurrency markets in terms of collective configurations, energetic coherence, and macroscopic order. The empirical analysis shows that the market alternates between regimes of strong collective alignment and more weakly organized states. Periods of large positive or negative magnetization correspond to ordered configurations in which most assets move coherently in the same direction, whereas values near zero indicate comparatively fragmented market behavior.

The joint energy--magnetization phase space reveals a clear structural relation between global alignment and interaction consistency. The empirical distribution is organized around an approximately parabolic relation between $E$ and $M$, showing that strongly aligned market states tend to occupy lower-energy regions. In addition, the magnetization distribution exhibits a pronounced bimodal structure, indicating that the market spends a substantial fraction of the observation period in highly ordered collective states rather than fluctuating primarily around neutral configurations. This finding suggests that synchronized market-wide behavior is not exceptional but instead constitutes a central organizing feature of the system. The phase-space distribution also displays an asymmetry between the positive and negative branches, indicating that coherent bearish regimes were more frequent or more persistent than comparable bullish regimes during the sample period.

Taken together, these results show that the Market Crystal provides a compact thermodynamic description of market organization, in which magnetization acts as an order parameter and the energy quantifies the compatibility of the instantaneous configuration with the empirical interaction network. More broadly, the framework offers a physically interpretable representation of collective behavior in financial systems and opens several directions for future research. Additional observables, such as susceptibility, heat capacity, or persistence measures, could be used to investigate possible market phase transitions and regime changes. Likewise, extending the model to incorporate time-dependent correlations, alternative embedding procedures, or multilayer interaction structures may provide deeper insight into the evolving architecture of cryptocurrency markets. In this sense, the Market Crystal establishes a systematic bridge between empirical financial data and the analytical framework of statistical mechanics.


\end{document}